\documentclass[12pt,preprint]{aastex}

\newcommand{\hst}{{\sl HST}}

\slugcomment{ApJ Letter, accepted on April 9$^{\rm th}$, 2008}
\shorttitle{NGC 6791}
\shortauthors{Bedin et al.}

\begin{document}

\def\subr #1{_{{\rm #1}}}


\title{The Puzzling White Dwarf Cooling Sequence in NGC 6791: \\ 
  A Simple Solution\altaffilmark{1}}

\altaffiltext{1}{ Based on observations  with the NASA/ESA {\it Hubble
Space Telescope},  obtained at the Space  Telescope Science Institute,
which is operated by AURA, Inc., under NASA contract NAS 5-26555.}

\author{ L.\ R.\ Bedin\altaffilmark{2},  
         M.\ Salaris\altaffilmark{3}, 
         G.\ Piotto\altaffilmark{4},  
         S.\ Cassisi\altaffilmark{5}, 
         A.\ P.\ Milone\altaffilmark{4}, \\
         J.\ Anderson\altaffilmark{2}, and 
         I.\ R.\ King\altaffilmark{6}.
         }

\altaffiltext{2}{Space Telescope Science Institute, 3800 San Martin 
Drive, Baltimore, MD 21218; [bedin; jayander]@stsci.edu}

\altaffiltext{3}{Astrophysics Research Institute, Liverpool John Moores
University, 12 Quays House, Birkenhead, CH41 1LD, UK; ms@astro.livjm.ac.uk}

\altaffiltext{4}{Dipartimento di Astronomia, Universit\`a di Padova,
Vicolo dell'Osservatorio 2, I-35122 Padova, Italy;
giampaolo.piotto@unipd.it}

\altaffiltext{5}{INAF-Osservatorio Astronomico di Collurania,
via M. Maggini, 64100 Teramo, Italy;
cassisi@oa-teramo.inaf.it}

\altaffiltext{6}{Department of Astronomy, University of Washington,
Box 351580, Seattle, WA 98195-1580; king@astro.washington.edu}

\begin{abstract}
In  this paper we  demonstrate that  the puzzling  bright peak  in the
luminosity function  of the white  dwarf (WD) cooling sequence  of NGC
6791 can be naturally accounted  for if $\sim34\%$ of the observed WDs
are WD+WD binary systems.
\end{abstract}


%
\section{Introduction}
%
 
NGC 6791  is a peculiar open cluster.   It is one of  the richest open
clusters, unusually old (8--9 Gyr,  Stetson et al.\ 2003, King et al.\
2005), and extremely  metal rich ([Fe/H] $\sim$ +0.4,  Gratton et al.\
2006, Carraro et al.\ 2006, Origlia et al.\ 2006).
NGC 6791  is close enough that  \hst/ACS imaging can  reach very faint
luminosities.  Because of this, it has been one of our targets for the
study  of the  bottom  of  the main  sequence  (proposals GO-9815  and
GO-10471,   PI:   King).    In   previous  papers   we   studied   the
color-magnitude diagram (CMD), mass function (King et al.\ 2005), and,
the white dwarf (WD) cooling sequence (CS).
The latter  investigation (Bedin et al.\  2005, Paper I,  and Bedin et
al.\  2008,  Paper  II)  provided  us  with  additional  exciting  and
unexpected results.

The most  important results are:  (i) The WD luminosity  function (LF)
shows  a peak  at a  magnitude (F606W  = $27.45\pm0.05$),  which would
imply a CO-core WD cooling age inconsistent by a factor $>2$\ with the
age  obtained  from the  MS  turn-off  (TO  --- Paper~I).   (ii)  With
second-epoch  HST observations,  in Paper~II  we identified  a second,
fainter peak in  the WD LF, as richly populated  as the brighter peak,
but at  a magnitude (F606W  = $28.15\pm0.05$) still  slightly brighter
than expected  from the  TO age.   (iii) In addition,  in Paper  II we
found that at the  magnitude level of each of the two  LF peaks the WD
cooling sequence describes  a sort of blue hook.  (iv)  The WD CS ends
at F606W $\sim$ 28.35.

As  discussed in  Paper~II,  the fainter  peak  in the  WD  LF may  be
consistent with  the expected position of  the bottom of the  CS for a
cluster with the  TO age of NGC  6791 if we account for  the effect of
$\rm ^{22}Ne$ diffusion  in the liquid phase (e.g.  Deloye \& Bildsten
2002,  Garc\'ia-Berro et  al.\ 2007)  plus smaller  contributions from
$\rm   ^{22}Ne$   separation    during   the   crystallization   phase
(Segretain~1996), plus uncertainties in the core CO profile due to the
$\rm ^{12}C(\alpha, \gamma)^{16}O$ reaction rate.

However, for the brighter peak in the WD LF, neither could we find any
convincing explanation, nor, even three years after the publication of
Paper~I, has a plausible scenario  to explain this anomaly appeared in
the literature.   Hansen (2005) and Kalirai et  al.\ (2007) speculated
that the  two peaks in  the CS LF  could be accounted for  by assuming
that the  cluster has two different  types of WDs. In  addition to the
usual CO-core WDs, they claim  that NGC 6791 also hosts an anomalously
high  number  of  very  massive   He-core  WDs  (masses  of  at  least
0.5$M_{\odot}$),  which produce the  observed bright  peak in  the LF.
The presence  of some  He-core WDs cannot  of course be  excluded, and
indeed Kalirai et al.\ (2007) have identified in the upper part of the
CS of NGC 6791 a few very  bright WDs whose masses (below the value of
the electron-degenerate core mass at the He flash) are consistent with
'normal' He-core  WDs.  The massive He-core WD  scenario, by contrast,
is  severely challenged  by theoretical  evidence, as  we exhaustively
discussed in Paper  II.  The main problem --- apart  from the need for
extremely  efficient mass  loss along  the red  giant branch  even for
masses well above  1$M_{\odot}$ --- is that the  mass of these He-core
WDs needs to be  sizably larger than the core mass at  the He flash in
metal-rich,  low-mass red-giant-branch stars,  which sets  the maximum
allowed mass for He-core WDs.  It is also not clear how to explain the
blue hook  of the CS at the  magnitude of the bright  peak, unless the
He-core WDs  reach masses much larger  than the already  high value of
0.5$M_{\odot}$.  In this paper we will show that no sizable population
of massive  He-core WDs is  needed to explain  the WD CS in  NGC 6791,
which may turn out to be less perplexing than originally thought.

%
\section{Binaries in NGC 6791}
\label{binary}
%

A conspicuous  feature of the CMD of  NGC 6791 is the  large number of
photometric binaries visible on the red  side of the MS (Fig.\ 1).  In
order  to  derive the  fraction  of  MS+MS  binaries with  mass  ratio
$q>0.5$, we applied  the recipes of Milone et  al.~(2008).  First, the
CMD was  cleaned of field stars  using the proper motions  of Bedin et
al.\ (2006).   We then  divided the  CMD in two  parts:\ a  region $A$
containing all the  single stars and the binaries  with a primary with
$17.5<m_{F814W}<21.0$ (shadowed  region in Fig.\  1) and a  region $B$
(darker region  in Fig.\ 1)  that is the  portion of $A$  populated by
binaries with $q>0.5$.  The bluest  line is the MS fiducial line moved
by  3$\sigma$ to  the blue,  where $\sigma$  is the  photometric error
coming  from  the artificial-star  tests  (Bedin  et  al.\ 2008).  The
reddest line  is the locus of  $q=1$ binaries, moved  3$\sigma$ to the
red.  The locus in the CMD of the binaries with a given mass ratio $q$
was found  using the mass-luminosity relation of  Pietrinferni et al.\
(2004).

\begin{figure}
\epsscale{1.00}
\plotone{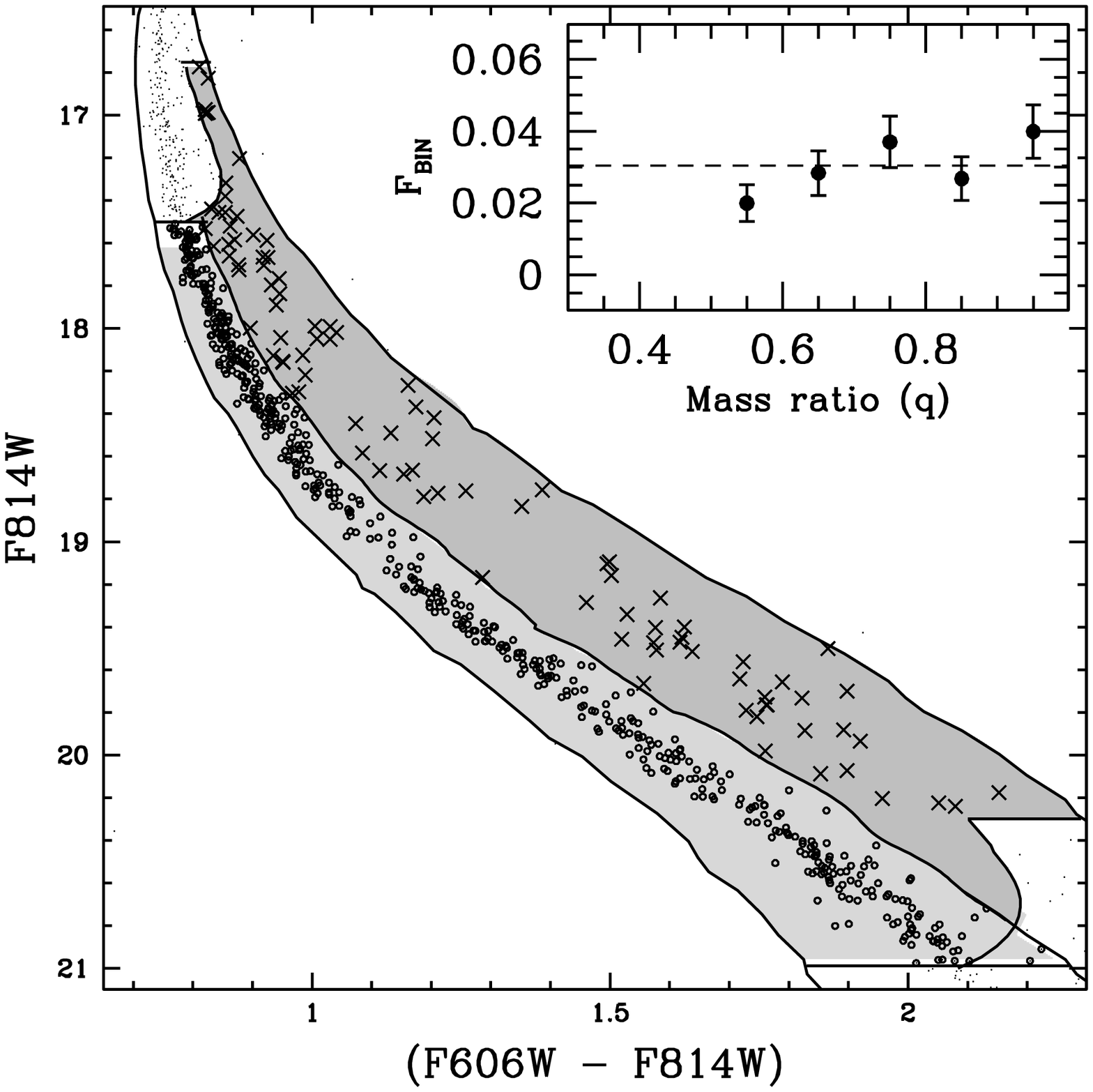}
\caption{The main  sequence of  NGC 6791 with  the binaries  with mass
fraction $q>0.5$ plotted as crosses. In the inset, the distribution of
the  binary fraction  for $q>0.5$.   Note that  we  used proper-motion
membership (Bedin  et al.\ 2006) to  remove field stars  from the CMD.
See text for the definitions of the shadowed regions.}
\label{MSMS}
\end{figure}

The  fraction  of  binaries  with  $q>0.5$ is  calculated  as  $f_{\rm
bin}^{q>0.5}  =  N_{\rm  OBS}^{\rm  B}/N_{\rm OBS}^{\rm  A}  -  N_{\rm
ART}^{\rm B}/N_{\rm  ART}^{\rm A}$, where $N_{\rm  OBS}^{\rm A(B)}$ is
the  number of  stars (corrected  for completeness),  observed  in the
region  $A$  ($B$); $N_{\rm  ART}^{\rm  A(B)}$  are the  corresponding
values  of  artificial stars.   (The  derivation  of  the equation  is
straightforward  but too  long  to include  here.)   We found  $f_{\rm
bin}^{q>0.5}= 0.16 \pm 0.02$.  We  repeated the same procedure for the
fractions of  binaries with $q>0.6$,  $0.7$, $0.8$ and $0.9$.  Then we
derived the  fractions of binaries  in five intervals of  size $\Delta
{q}=0.1$ in the range  $0.5<q\leq1$. [E.g., we calculated the fraction
of binaries  with $0.5<q\leq0.6$: $f_{\rm  bin}^{0.5<q<0.6}$= $(f_{\rm
bin}^{q>0.5}-  f_{\rm  bin}^{q>0.6})$,  and  similarly for  the  other
bins.]  The results are plotted  in the inset of Fig.\ \ref{MSMS}. The
distribution  of  the  binary  fraction   as  a  function  of  $q$  is
flat.  Extrapolating  this  flat  distribution to  lower  mass  ratios
($q<0.5$), we  estimate that  the total MS+MS  binary fraction  in the
core of NGC~6791 is $f_{\rm bin}^{\rm tot}\simeq 32 \pm 3 \%$.

We have  also calculated the  global fraction of binaries  using other
methods described  in the literature,  and found similar  results.  In
particular, we also estimated  the binary fraction using the procedure
described by  Sollima et al.\ (2007,  see their Section\  5). The only
differences from  what described  in that paper  are that we  used the
mass-luminosity relation  of Pietrinferni et al.\ (2004),  and that we
removed  the field  stars using  proper  motions (and  not a  Galactic
model, as  in Sollima et  al.)  The Sollima  et al.\ (2007)  method is
based  on simulations  of the  photometric binary  population assuming
different mass-ratio  distributions, followed by  a comparison between
the observed and simulated CMDs (see their paper for further details).
If we assume the binary fraction  to be the function of the mass ratio
that  Fisher   et  al.\  (2005)   found  for  the   solar  neighborood
distribution, we find for NGC 6791 $f_{\rm bin}^{\rm tot}=25\%$; if we
get our distribution by randomly  coupling stars from the De Marchi et
al.\  (2005)   initial  mass   function,  we  get   $f_{\rm  bin}^{\rm
tot}=37\%$.

Photometric  binaries are  not the  only evidence  for a  large binary
population  in  NGC  6791.   In  a  recent  large  survey  of  stellar
variability in a field  of $\sim30\times30$ arcmin$^2$ centered on the
center of NGC  6791, De Marchi et al.\ (2007)  have identified a large
number of  variable stars whose  variability is associated  with their
binary nature.  In particular, De  Marchi et al.\ brought to three the
number of known  cataclysmic variables (CV) in NGC  6791, showing that
on the  basis of  their proper motions  they are  all high-probability
cluster  members.  In  addition, De  Marchi et  al.\ found  29 contact
binaries and 61 eclipsing systems.

A precise determination of the total number of binaries in NGC 6791 is
beyond the  purpose of  the present paper.  However, it is  clear that
this cluster hosts a  large binary population (surely $>$25--30\%), as
it is  clear that photometric  binaries are only  one of the  kinds of
binary  present.  It is  also evident  from the  many CVs,  that white
dwarfs are  commonly found  in binaries, as  well.  Moreover,  we note
that in the  cluster core, where the  ACS field of Papers I  and II is
centered,  the  fraction  of  binaries  with  massive  components  (of
interest for the  discussion in the following section)  must be higher
than the average fraction  in the cluster, because of mass-segregation
effects.

%
\section{The effect of binaries on the WD CS}
\label{theory}
%

In the  previous section we have  shown that NGC 6791  hosts a sizable
fraction of stars in binary  systems. Consequently, we can expect that
a high fraction  of NGC 6791 WDs  to be in binaries, and  part of them
must  be  WD+WD  binary  systems.   Prompted  by  this  reasoning,  we
investigated the effect of double WD binaries on the WD CS.

To  this purpose, we  simulated the  WD cooling  sequence of  NGC 6791
using the same  WD isochrones described in Paper~II,  for a progenitor
metallicity [Fe/H] = +0.4.   Each isochrone provides the magnitudes of
a single-age,  single-metallicity CS  in the ACS/WFC  Vega-mag system,
together with the  mass of the evolving WD at each  point along the CS
and the  corresponding progenitor mass.  We used  an apparent distance
modulus    $(m-M)_{\rm   F606W}=13.44$    and   $E(F606W-F814W)=0.14$,
consistent  with   Papers  I  and  II.   We   adopted  as  theoretical
counterpart  of  the single  WDs  in  the  observed CS  the  6-Gyr-old
isochrone  that matches  the  fainter peak  of  the LF.   This age  is
younger than the TO age.  As briefly outlined in the Introduction, and
discussed in  detail in Paper II,  this discrepancy may  be removed if
some new  developments in the physics  of WD cooling  (which will slow
down the cooling process) are implemented in CO-core WD modeling.

Theoretical cooling  sequences that  include a fraction  of unresolved
non-interacting  WD$+$WD binaries  have been  computed by  means  of a
Monte-Carlo  (MC) simulation.  A value  of the  progenitor mass  for a
generic single  WD along the CS  is extracted randomly  according to a
Salpeter initial mass function (IMF). The mass and the F606W and F814W
magnitudes of  the WD are  then determined by  quadratic interpolation
along  the isochrone  points.  To  assign a  companion to  this  WD we
extract randomly a  value of $q$ according to  the adopted statistical
distribution (see below), and the corresponding WD mass and magnitudes
are calculated using  the isochrone. The fluxes of  the two components
in the F606W and F814W filters are added, and the total magnitudes and
colors  of the  composite  system  are computed.   We  then added  the
distance modulus  to these magnitudes, and perturbed  them randomly by
using  a  Gaussian  photometric  error  with  the  $\sigma$  from  the
artificial-star  tests. A theoretical  LF was  then computed  for this
synthetic population, and compared with the observed one.

\begin{figure}
\epsscale{1.00}
\plotone{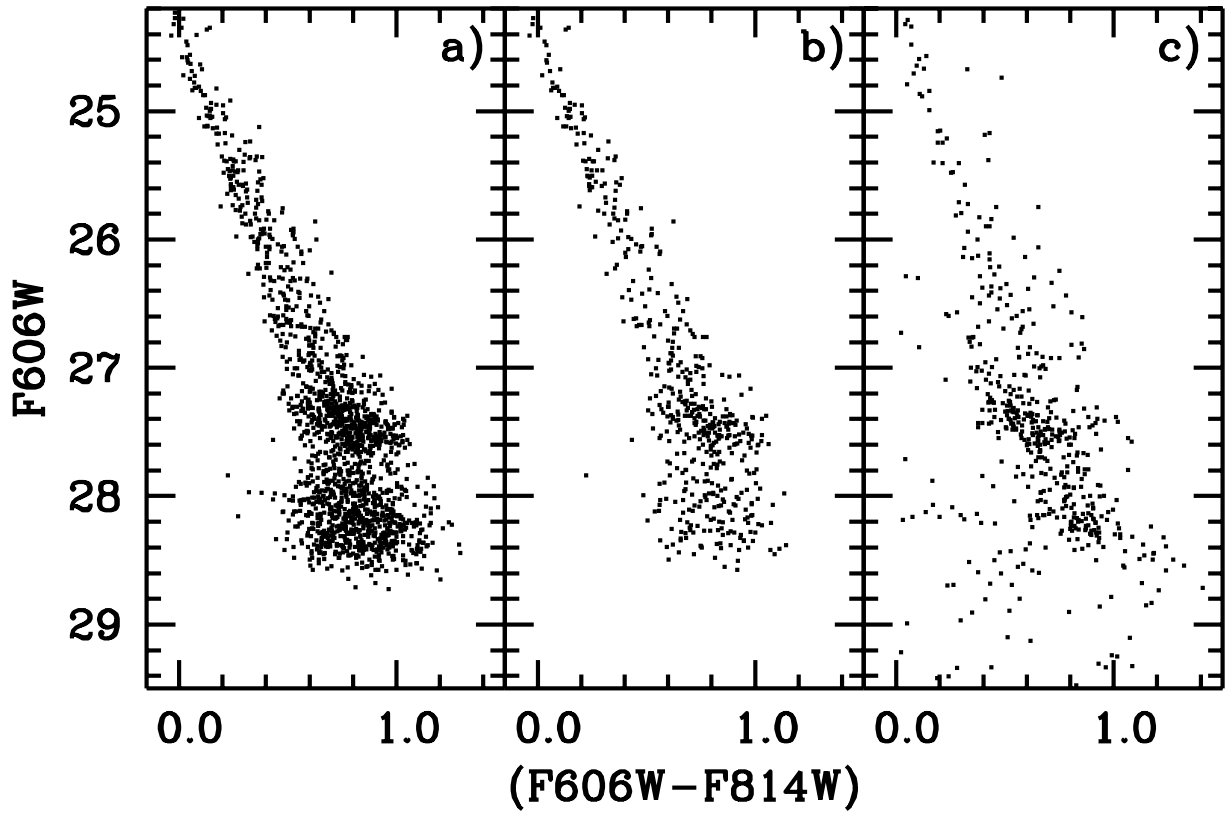}
\caption{Comparison  between observed  (panel {\bf  c})  and simulated
(panels {\bf a} and {\bf b}) CMDs of the WD CS.
Panel {\bf a}  shows the results of one MC  simulation.  Panel {\bf b}
shows  the same  simulation as  in {\bf  a}, but  with  stars randomly
removed in order  to account for the incompleteness,  as obtained from
artificial star tests (see Paper  II). The resulting number of objects
along  the CS is  approximately the  same as  in the  observed diagram
(panel  {\bf c}).   The synthetic  CMDs contain  34\% of  WD+WD binary
systems.  The  faint objects at  F606W$ < 28.4$ are  mostly background
galaxies,  as shown  in  Fig.\ 9  of  Bedin et  al.\  (2008). At  that
magnitude  level we  could not  use  proper motion  to select  cluster
members.}
\label{cmdC}
\end{figure}

\begin{figure}
\epsscale{1.00}
\plotone{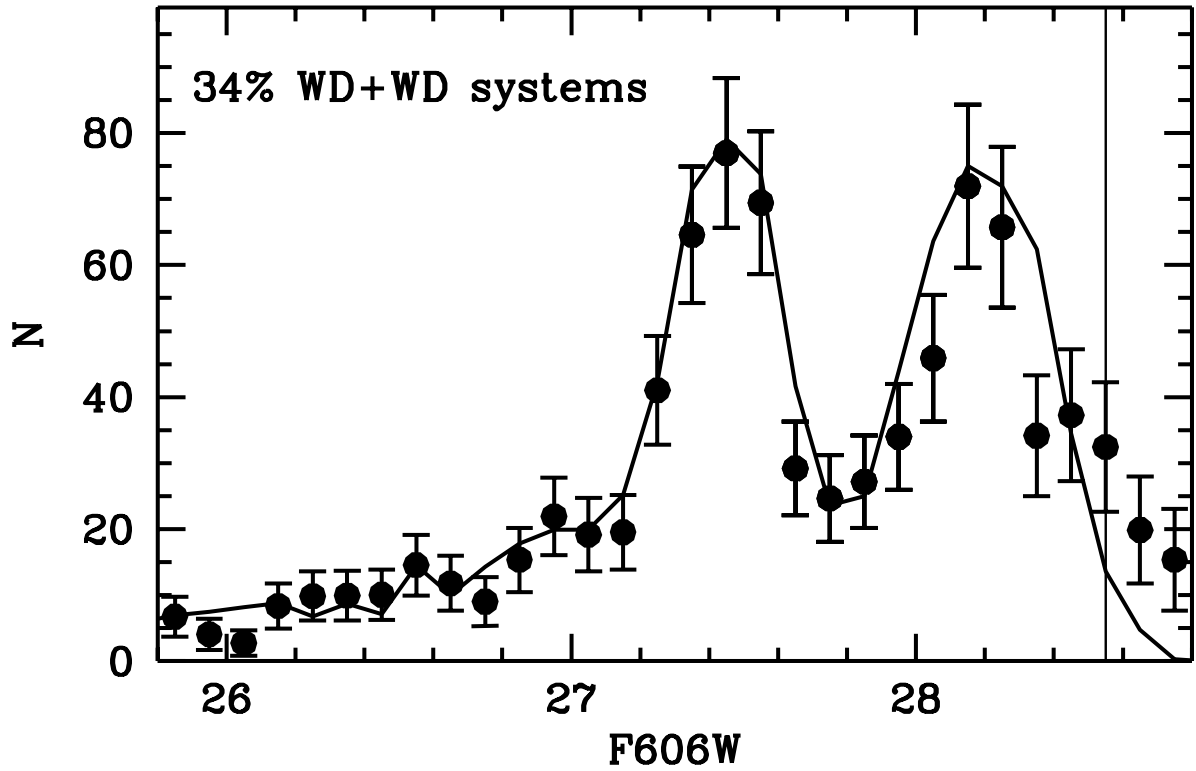}
\caption{Comparison  between  observed  (points with  error-bars)  and
simulated, completeness-corrected  WD LF assuming a  fraction of WD+WD
binary  systems equal  to 34\%.  A vertical  line shows  the  limit of
reliability of our completeness correction.  }
\label{wdlfC}
\end{figure}

Figures \ref{cmdC} (panels  {\it a} and {\it b})  and \ref{wdlfC} show
the CMDs and the LF  of a synthetic cooling sequence computed assuming
a fraction of binary systems of 54\%  and a mass ratio $q$ with a flat
distribution between 0.5 and 1.0,  as obtained for the MS+MS binaries,
see  inset in  Fig.~1.  (The  effect of  a different  distribution for
$q<0.5$ will  be briefly discussed below.)  The  54\% overall fraction
leads to a  $\sim34$\% of WD+WD binaries on the  CS. This lower number
is a  consequence of the  fact that the  companion of a star  that has
become a WD may have a mass too low to have evolved to the WD stage at
the present age of the cluster.

The simulation  of the LF  includes a factor  of 10 more WDs  than the
observed  ones  ---  in  order   to  minimize  the  effect  of  random
fluctuations in  a given  magnitude bin ---  and has been  rescaled to
match the number (allowing for the completeness correction) of objects
in the  magnitude interval $26.0<$  F606W $<26.8$.  A  comparison with
the  observed CMD  (Fig.~\ref{cmdC},  right panel)  and  the F606W  LF
(Fig.~\ref{wdlfC}) shows  an astonishing overall  agreement. The WD+WD
pairs  produce a secondary,  brighter peak  in the  LF that  agrees in
magnitude, width,  and height with the observed  one.  The explanation
for this additional peak that  matches the observations so well can be
found by examining the corresponding CMD.

The synthetic CMD displays a  clump of stars at approximately constant
F606W  brightness, extended  towards the  blue, at  the bottom  of the
sequence, corresponding to the end of the single WD CS. In the case of
a single WD population,  this clump contains the overwhelming majority
of WDs.

The WD masses that  populate the clump range from $\sim$0.6$M_{\odot}$
to  $\sim$1$M_{\odot}$, and  the corresponding  progenitor  masses are
between  $\sim$1.8--2.0$M_{\odot}$   and  $\sim$5--6$M_{\odot}$.   For
comparison,   the  progenitor   mass  at   the  top   of  the   CS  is
$\sim$1.3$M_{\odot}$,    and   the    corresponding    WD   mass    is
$\sim$0.56$M_{\odot}$.  Given that the WDs  in the clump at the bottom
of the  CS span almost the  full range of progenitor  masses, the most
probable outcome for a WD+WD  system is obtained by combining the flux
of two WDs randomly selected  from this clump.  As the F606W magnitude
is  approximately constant  along the  clump, we  will have  a similar
feature in the CMD, but $\sim0.75$ magnitude brighter.  This naturally
explains the  appearance of an  additional brighter clump in  the CMD,
why it  turns to  the blue,  and also the  bright peak  in the  LF, at
exactly the location observed.

The prediction of a secondary bright peak (and its magnitude location)
in  the  LF  produced  by  binary  WD+WD  systems  is  robust  against
reasonable trends of the mass ratio $q$ of their progenitors.  We made
simulations using a flat $q$ distribution from zero to unity.
We  also  tested  the  possibility  that  the  $q$  distributions  are
single-valued,  and  we experimented  with  different  values of  $q$,
between 0.5 and 1.0.  The only major effect is that the total fraction
of binaries must be changed for varying $q$ distributions, in order to
match the height  of the observed bright peak in the  LF. On the other
hand, the fraction of WD+WD systems  needed to match the height of the
peak is only marginally affected.

A few WD+WD systems with progenitors below 1.6--1.8$M_{\odot}$ must be
present also in the upper CS (above the brighter peak). Their presence
makes the CS wider in color  at fixed $F606W$, compared to the case of
single WDs.  To  further test our binary scenario  --- as suggested by
the referee --- we have  considered the upper part ($F606W\le$26.5) of
the theoretical and observed sequences in panels b) and c) of Fig.\ 2.
We  subtracted  from  the  $(F606W-F814W)$ colors  of  the  individual
objects the corresponding color of  the fiducial WD CS obtained from a
linear  fit   to  the   CMD  of  the   simulated  and   observed  CSs,
respectively. In  this way,  both sequences are  made vertical  in the
CMD, with intrinsic color widths that can now be easily compared.  For
$F606W\le$26.5, the  rectified synthetic sequence  has a color  rms of
0.07  magnitude  in $(F606W-F814W)$,  smaller  than  the  rms of  0.13
magnitude of the observed CS.  Similar results
are found for  25.2 $ < F606W  < $ 25.8 (0.06 and  0.11 magnitude) and
for 26.0  $ < F606W < $  26.5 (0.09 and 0.17  magnitude).  In summary,
simulated WD+WD  binaries do not produce  a CS that is  wider in color
than  observed.  The broader  observed CS  is very  likely due  to the
presence of some  low mass He-core WDs, which must  be redder than the
CO-core WDs (see Paper I, II).

Another clear evidence of the binary nature of the brighter peak would
be  its concentration to  the center  of the  cluster.  Frustratingly,
however, this is  out of our reach, because  our field lies completely
within the  core of the cluster,  where the density  should be roughly
constant.  (An  unpublished star count by  one of us  [I.R.K.] gives a
core radius  of about 2.5 arcmin  --- a little more  than the distance
from the center of our field to its corner.)

%
\section{CONCLUSIONS}
%

In this paper we have demonstrated that the anomalously bright peak in
the luminosity  function of the  white dwarf (WD) cooling  sequence of
NGC 6791 discovered by Bedin et al.\ (2005) can be naturally accounted
for  if $\sim34$\%  of  the  observed WDs  are  actually WD+WD  binary
systems.

This population of double WDs  requires that about 50\% of the objects
in  NGC 6791  be  binaries. Such  a  fraction of  binaries is  totally
plausible for an open cluster  like NGC 6791. We demonstrated that NGC
6791  has a  fraction  of MS+MS  binaries  of the  order of  25--35\%,
similar to the fraction of  photometric binaries found in M67, another
old,  massive open  cluster (Montgomery  et al.\  1993).  As  shown by
Hurley  et al.\  (2005), a  model which  takes into  full  account the
dynamical  properties  of  M67  and  the  properties  of  its  stellar
population  requires  a  fraction  of   the  order  of  60\%  for  the
present-day binaries  (starting from  a primordial binary  fraction of
50\%), values very  similar to those required to explain  the WD CS of
NGC~6791.   The WD LF  down to the  bottom of the  CS in M67  has been
studied by Richer et al.\ (1998). Unfortunately, the large photometric
errors, the small  number of measured WDs, and  the consequently large
LF bin size  (0.5 magnitude) do not allow  investigation of the effect
of the large binary population of M67 on the WD CS.

We can anticipate that a secondary peak, about 0.75 magnitude brighter
than the clump at  the bottom of the CS, is also  present in the WD CS
of M4 (Bedin  et al., in preparation). The  presence of WD+WD binaries
in M4 has also been suggested by Hansen et al. (2004), who showed that
a binary fraction up to 10\%  does not affect the M4 age inferred from
the WDCS.

As  a final  note,  we want  to  emphasize that  a  large fraction  of
binaries implies  also the presence of many  interacting binaries. The
three CVs  and the  many contact binaries  discovered by De  Marchi et
al. (2007) provide observational support to this fact. We suspect that
this high binary  fraction may be related to  the other peculiarity of
this cluster,  the presence  of a blue  horizontal branch  (Kaluzny \&
Udalski  1992,   Stetson  et  al.\  2003),   despite  its  super-solar
metallicity.    Interestingly  enough,  the   presence  of   close  or
interacting  binaries with  the consequent  mass-loss  enhancement can
also   explain  the  presence   of  some   He-core  WDs,   such  those
spectroscopically identified  by Kalirai et  al.\ (2007) in  the upper
part of the CS.

\acknowledgements
We thank Brad Hansen, one of  the referees, for his careful reading of
the manuscript and for useful suggestions.
J.A.\ and  I.R.K.\ acknowledge support  from STScI grants  GO-9815 and
GO-10471. G.P.\ and A.M.\ acknowledge partial support from the Agenzia
Spaziale Italiana under contract ASI/088/06/0.

\clearpage

\end{document}